\documentclass[onecolumn,showpacs,amsmath,amssymb]{revtex4-1}
\usepackage{graphicx}
\newcommand{\beao}{\begin{eqnarray*}}\newcommand{\eeao}{\end{eqnarray*}}
\newcommand{\be}{\begin{equation}}\newcommand{\ee}{\end{equation}}
\newcommand{\bwe}{\begin{widetext}\begin{equation}}
\newcommand{\ewe}{\end{equation}\end{widetext}}
\newcommand{\bea}{\begin{eqnarray}}\newcommand{\eea}{\end{eqnarray}}
\newcommand{\beq}{\begin{eqnarray}}\newcommand{\eeq}{\end{eqnarray}}
\newcommand{\nn}{\nonumber}\newcommand{\pa}{\partial}
\newcommand{\ep}{\varepsilon}

\newcommand{\Ref}[1]{(\ref{#1})}
\newcommand{\F}{{\cal F}}
\newcommand{\Tr}{{\rm Tr}}
\begin{document}
\thispagestyle{empty}
\title{Vacuum energy between a sphere and a plane at finite temperature}

\author{M.~Bordag\footnote{bordag@itp.uni-leipzig.de}}
\author{I.~Pirozhenko\footnote{pirozhen@theor.jinr.ru}}

\affiliation{Leipzig University, Vor dem Hospitaltore 1, D-04103
Leipzig,
Germany  \\ and\\
Bogoliubov Laboratory of Theoretical Physics, Joint Institute for
Nuclear Research, Dubna 141 980, Russia}
\begin{abstract}
We consider the Casimir effect for a sphere in front of a plane at
finite temperature for scalar and electromagnetic fields and
calculate the limiting cases. For small separation we compare the
exact results with the corresponding ones obtained in proximity
force approximation. For the scalar field with Dirichlet boundary
conditions, the low temperature correction is of order $T^2$ like
for parallel planes. For the electromagnetic field it is of order
$T^4$. For high temperature we observe the usual picture that the
leading order is given by the zeroth Matsubara frequency. The
non-zero frequencies are exponentially suppressed except for the
case of close separation.
\end{abstract}
\maketitle
\section{Introduction}
By now, the Casimir force between a sphere and a plane was
calculated using the functional determinant (or TGTG- or
T-matrix-) representation and, for scalar field, numerically by
the world line methods, and, of course, in the Proximity Force
Approximation (PFA). The extension of these methods to finite
temperature is in its very beginning. Even for the PFA we did not
find in literature a representation at finite  temperature,
although the calculation is quite simple. Recently
\cite{Weber:2009dp,Gies:2009nn}, using the world line method, the
temperature dependence was studied and an interesting interplay
between geometry and temperature was observed. Using the
functional determinant representation in \cite{Durand2009} a
partly numerical, partly analytical study at medium and large
separation was represented where also dielectric materials were
included. It must be mentioned that the corrections beyond PFA are
a topic of actual interest also for high precision measurements of
the Casimir force where it is necessary to know the temperature
corrections at small separation.

In this paper we calculate the free energy at finite temperature
for a sphere in front of a plane for conductor boundary conditions
using the functional determinant method which gives the exact
result. We pay special attention to the case of small separation
and to the limiting cases of low and high temperature. Also, for
small separation we establish the relation to the Proximity Force
Approximation (PFA) and determine where it is applicable.

The paper is organized as follows. In the next section we prepare
the necessary formulas of the functional determinant
representation. In sections 3 and 4 we calculate the temperature
dependent part of the free energy at small separation by both
methods, PFA and the functional determinant method. In sections 5
and 6 we derive the low and high temperature expansions using the
functional determinant method. Finally, the results are discussed
in section 7.

\noindent Throughout the paper we use units with $\hbar=c=k_{\rm
B}=1$.

\section{Basic formulas}
In this section we collect the formulas we need for the free energy in
sphere-plane geometry.
For the free energy we use the standard Matsubara-representation.
For the interaction energy between a sphere and a plane we use
it together with the
functional determinant  representation,
\be\label{1.F2} \F=\frac{T}{2}\sum_{n=-\infty}^{\infty} \Tr \ln
\left(1-\mathbf{M}(\xi_n)\right),
\ee
where $\xi_n=2\pi Tn$ are the Matsubara frequencies. At zero temperature,
the Matsubara sum becomes an integration,
$T\sum_{n=-\infty}^{\infty}\to \int_{-\infty}^\infty\frac{d\xi}{2\pi}$
with $\xi_n\to\xi$ and
the free energy turns into the vacuum energy,
\be\label{1.E}
E_0=\frac{1}{2}\int_{-\infty}^\infty\frac{d\xi}{2\pi}\
\Tr \ln \left(1-\mathbf{M}(\xi)\right).
\ee
In these formulas,  $\mathbf{M}(\xi)$
is a matrix in the orbital momentum indices $l$ and $l'$,
\be\label{1.M} M^{\rm  }_{l,l'}(\xi) =d^{\rm }_l(\xi R)
\sqrt{\frac{\pi}{4\xi L}}\sum_{l''=|l-l'|}^{l+l'}K_{\nu''}(2\xi
L)\,H_{ll'}^{l''}\,
 \,.
\ee
For Dirichlet boundary conditions on the sphere,
\be\label{1.d}
d^{\rm  }_l(\xi )=\frac{I_{\nu'}(\xi)}{K_{\nu}(\xi )}
\ee
results from the scattering T-matrix. In these formulas,
$I_{l+1/2}(x)$ and $K_{l+1/2}(x)$ are the modified Bessel
functions. The geometry is shown in Fig. 1.
We introduced the notations $\nu=l+1/2$, $\nu'=l'+1/2$ and $\nu''=l''+1/2$,
which will be used throughout the paper.  \\
\begin{figure}[h]
\includegraphics[width=7cm]{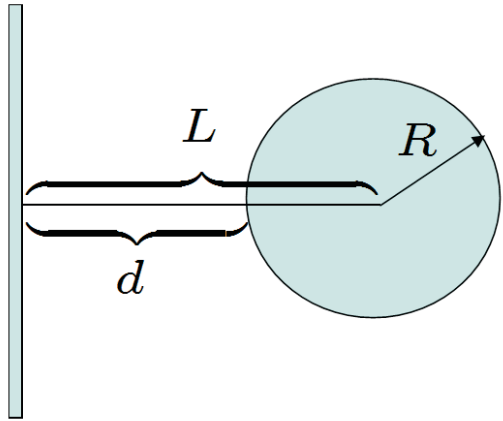}
\caption{The configuration of a sphere in front of a plane.}\end{figure}
The above formulas can be found in the original papers on the
functional determinant method \cite{WIRZBA2006},
\cite{Emig:2007cf}. Here we follow the notations used in
\cite{BORDAG2008C} and in \cite{BKMM}. The factors $H_{ll'}^{l''}$
in \Ref{1.M} result from the translation formulas. Their explicit
form is
\bea\label{1.H} 
H_{ll'}^{l''}&=&  \sqrt{(2l+1)(2l'+1)}(2l''+1) \nn\\&&\times
\left(\begin{array}{ccc}l&l'&l''\\0&0&0\end{array}\right)
\left(\begin{array}{ccc}l&l'&l''\\m&-m&0\end{array}\right), \eea
where the parentheses denote the $3j$-symbols. For Neumann
boundary conditions on the sphere we have to substitute $d_l(x)$
by
\be\label{1.N}
d^{\rm N}_l(\xi )=
\frac{\left(I_{\nu'}(\xi)/\sqrt{\xi }\right)'}{\left(K_{\nu}(\xi )/\sqrt{\xi}\right)'}
\ee
and for Neumann boundary conditions on the plane we have to
reverse the sign in the logarithm in \Ref{1.F2}.

For the electromagnetic field the matrix $\mathbf{M}$ acquires
indices for the two polarizations, which correspond to the TE and
the TM modes in spherical geometry,
\bea\label{1.NED} \mathbb{M}_{l,l'}
&=&
\sqrt{\frac{\pi}{4\xi L}}\sum_{l''=|l-l'|}^{l+l'}K_{\nu''}(2\xi
L) H_{ll'}^{l''}\,
\\&&\nn
\times\left(\begin{array}{cc}\Lambda_{l,l'}^{l''}&\tilde{\Lambda}_{l,l'}
\\ \tilde{\Lambda}_{l,l'}&\Lambda_{l,l'}^{l''}\end{array}\right)
\left(\begin{array}{cc}d^{\rm TE}_l(\xi R)&0
\\ 0&-d^{\rm TM}_l(\xi R)\end{array}\right)
 \,.
\eea
with the factors
\bea\label{2.LA}
\Lambda_{ll'}^{l''}&=&\frac{\frac12\left[l''(l''+1)-l(l+1)-l'(l'+1)\right]}
{\sqrt{l(l+1)l'(l'+1)}}\,,
\nn\\
\tilde{\Lambda}_{ll'}&=&\frac{2m\xi L}{\sqrt{l(l+1)l'(l'+1)}}
\eea
which follow from the translation formulas for the vector field. The factors resulting
from
the scattering T-matrices are
\be\label{1.dTE} d^{\rm TE}_l(\xi R)= \frac{I_{\nu'}(\xi
R)}{K_{\nu}(\xi R)}\,,
\ee
which is the same as for the scalar field with Dirichlet boundary
conditions, and, for the TM-mode,
\be\label{1.dTM} d^{\rm TM}_l(\xi R)= \frac{\left(I_{\nu'}(\xi
R)\sqrt{\xi R}\right)'} {\left(K_{\nu}(\xi R)\sqrt{\xi
R}\right)'}\,.
\ee
When inserting these expressions into \Ref{1.F2} or \Ref{1.E}, the
trace must be taken also over the polarizations. More details are
given in \cite{BORDAG2009C}.

\section{The PFA at finite temperature}
In order to discuss the free energy at small separation which is
the relevant scale for PFA
 we first define the corresponding temperature
regions of interest. These are
\be \label{4.tr}
\begin{array}{lrrrrr}
\mbox{1. low temperature:}     & dT&\ll &RT &\ll &1  \,,  \\
\mbox{2. medium temperature:~~~~}  & dT&\ll &1&\ll & RT   \,, \\
\mbox{3. high temperature:}  & 1 &\ll& dT &\ll& RT \,.
\end{array}
\ee
In each case $d\ll R$ holds. Further we introduce
\be\label{2.d}
                   \ep=\frac{d}{R}
\ee
as small parameter.

Below we consider the  PFA at finite temperature in some more
detail as in   \cite{Bordag:2001qi}, section 5.1.2. We do this for
the electromagnetic field with conductor boundary conditions.
 For that,
first we need to remember the free energy $\F_{||}$ per unit area
for two parallel planes. It can be obtained in several ways, for
instance as special case from the general formula \Ref{1.F2}
\be\label{Fp} \F_{||}(d,T)=
    T\sum_{n=-\infty}^\infty\,
    \int\frac{d \mathbf{k}}{(2\pi)^2}\,
    \ln\left(1-e^{-2d\sqrt{\xi_n^2+\mathbf{k}^2}}\,\right),
\ee
where $\mathbf{k}$ is the momentum in parallel to the planes
and $\xi_n$ are the Matsubara frequencies.
By representing this free energy in the form
\be\label{4.F} \F_{||}(d,T)=-\frac{\pi^2}{720 d^3}\, (1+g(2dT))\,
\ee
%
($2d=1/T_{\rm eff}$ is the inverse effective temperature)
we defined
the function $g(x)$ which describes the temperature dependence.
From \Ref{Fp} it can be written in the form
\be\label{g1}
    g(x)=-1+\frac{45\zeta(3)x}{\pi^3}
        -\frac{90x}{\pi^3}\sum_{n=1}^{\infty}
        \int_0^\infty dk\,k\,\ln\left(1-e^{-\sqrt{(2\pi n x)^2+k^2}}\,\right).
\ee
A sum representation  can be obtained from \Ref{Fp} by expanding
the logarithm into a series with subsequent integration over k and
summation over n,
\be\label{g2}
    g(x)=-1+45 x^4\sum_{m=1}^\infty
    \left[  \frac{\coth(m\pi  x)}{(m\pi x)^3}
            +\left(m\pi x\,\sinh(m\pi  x)\right)^{-2}
    \right]\,.
\ee
Another representation can be obtained from \Ref{g1} using the
Abel-Plana formula, see eq. (7.81) in \cite{BKMM} for example.
Comparing that representation with \Ref{g2}   the inversion
symmetry
\be\label{inversion}g(x)=x^4\,g\left(\frac{1}{x}\right),  \ee
which holds for the function $g(x)$, can be established.
The asymptotic expansions of this function for small and large
arguments are
\be\label{g3} g(x)\sim    \left\{
        \begin{array}{c}\frac{45\zeta(3)}{\pi^3}\,x^3-x^4+\dots \quad\mbox{for $x\to0$,}\\[5pt]
        \frac{45\zeta(3)}{\pi^3}\,x -1+\dots \quad\mbox{for $x\to\infty$.}
        \end{array}  \right.
\ee
The upper line follows directly from \Ref{g2} and the lower line
from \Ref{inversion}. In \Ref{g3}, the dots denote exponentially
small contributions. The asymptotic formulas reproduce $g(x)$
quite accurate everywhere. Even at the symmetry point $x=1$
 the relative deviation is less than $5\%$.

%
%

Now we apply the idea of PFA to \Ref{4.F}.  We have to integrate
$\F_{||}(d,T)$ over the area of a circle of radius $R$ in the
$(x,y)$-plane,
\be\label{4.FP1}\F^{\rm PFA}=
\int dx dy\, \F_{||}(d+h(x,y),T),
\ee
where $d+h(x,y)$ is the separation between the plane and the
sphere at the point $(x,y)$ in the plane. Using polar coordinates
we get
\be\label{4.FP1a}
\F^{\rm PFA}=   \int_0^Rdr\, r\,\int_0^{2\pi}d\varphi \, \F_{||}(d+R(1-\cos\theta),T)
\ee
with $r=R\sin\theta$. Using the azimuthal symmetry and changing
for the variable $t=1-\cos\theta$ we come to the representation
\be\label{4.FP2} \F^{\rm PFA}=    2\pi R^2\int_0^1dt\,(1-t)\,
\F_{||}(d+Rt,T)\,. \ee
The corresponding approximation for the force,
\be\label{4.f1}f^{\rm PFA}=-\frac{\pa}{\pa d}\,\F^{\rm PFA}\,,
\ee
can be written in the form
\be\label{4.f2}f^{\rm PFA}=2\pi R\left(\F_{||}(d,T)-\int_0^1dt\,\F_{||}(d+Rt,T)\right)\,.
\ee
In deriving \Ref{4.f2}, for the derivative of $\F_{||}$ we used
$\pa/\pa d=(1/R)\pa/\pa t$ and integrated by parts.

Both expressions, \Ref{4.FP2} and \Ref{4.f2}, are meaningful for
$\ep\to0$ only. However, further simplification depends on the specific properties of
the functions involved.

The simplest case is   $T=0$. Here the function $ \F_{||}(d,0)$ is just
the vacuum energy
\be\label{E0}
E_0=-\frac{\pi^3}{720 d^3}.\ee
It increases proportional to $1/d^3$ with
decreasing separation $d$. In the limit $\ep\to0$, the second term
in the parenthesis in \Ref{4.f2} does not contribute  and one
arrives at the simple rule
\be\label{rule}f^{\rm PFA}_{|_{{T=0}}}=2\pi R\, \F_{||}(d,0) \ee
called {\it proximity force theorem} in \cite{Blocki}.

For a general function $\F_{||}(d,T)$, the validity of \Ref{rule}
depends on the behavior $\F_{||}(d,T)\sim d^{-\alpha}$ for
$d\to0$. From Eq.\Ref{4.f2} it is easy to show the following. If
$\alpha>0$, i.e., if $\F_{||}(d,T)$ increases for decreasing
separation, the integral term in \Ref{4.f2} is by a factor $d$ (or
$d\ln d$ for $\alpha=1$) smaller than the first term in the
parenthesis in \Ref{4.f2} and \Ref{rule} holds. Contrary, for
$\alpha<0$ the integral term is dominating and \Ref{rule} does not
hold. A constant contribution to $\F_{||}(d,T)$, having $\alpha
=0$, does not contribute to the force and \Ref{rule} does not hold
either.

Now we consider  \Ref{4.FP2} and  \Ref{4.f2}  with the free energy
\Ref{4.F} of parallel planes inserted. We get
\be\label{F1}
                  \F^{\rm PFA}=-\frac{\pi^3 R}{720 d^2}\left[
                  \frac{1}{1+\ep}+2\ep^2\int_0^1dt\,(1-t)\,\frac{g(2(\ep+t)RT)}{(\ep+t)^3}
                  \right]\,
\ee
for the free energy and
\be\label{f1} f^{\rm PFA}=-\frac{\pi^3 R}{360 d^3}\left[
                  \frac{1+3/2\,\ep}{(1+\ep)^2}+g(2dT)-\ep^3\int_0^1dt\,\frac{g(2(\ep+t)RT)}{(\ep+t)^3}
                  \right]\,
\ee
for the force. These expressions are meaningful in the limit of
$\ep\to0$ only. First of all we mention that the corrections to
the vacuum  energy in the first terms in the square brackets in
both expressions, which are  due to the geometry,  are known to be
beyond the precision of PFA and we drop them in the following. In
order to perform this limit in the temperature dependent parts it
is meaningful to consider two cases.

First we consider low and medium temperature as defined in
\Ref{4.tr}. For this we substitute
\be \label{lm} dT=\ep\, RT \ee
in \Ref{F1} and \Ref{f1} and consider the limit $\ep\to0$ with
$RT$ fixed. In the integrals we can put $\ep=0$ directly and using
\Ref{g3} we obtain
\be\label{F2}
                  \F^{\rm PFA}=-\frac{\pi^3 R}{720 d^2}\left[
                  1+2\ep^2\int_0^1dt\,(1-t)\,\frac{g(2tRT)}{t^3}+\dots
                  \right]\,
\ee
for the free energy and
\be\label{f2} f^{\rm PFA}=-\frac{\pi^3 R}{360 d^3}\left[
                 1+\ep^3\frac{360\zeta(3)}{\pi^3}(RT)^3-\ep^3\int_0^1dt\,\frac{g(2tRT)}{t^3}
                  +\dots\right]\,
\ee
for the force. These expansions are uniform in $RT$.

We note that the temperature correction to the free energy does
not depend on the separation $d$ and that only the next order
contribution not shown in \Ref{F2} gives the contribution to the
force which is shown in \Ref{f2}. In opposite to \Ref{g3},  in the
above formulas the dots denote contributions which are suppressed
only by powers of the small parameter.

As special cases, from \Ref{F2} and \Ref{f2}, we obtain for
$dT\ll1$, i.e., for low temperature,
\be\label{F3}
                  \F^{\rm PFA}=-\frac{\pi^3 R}{720 d^2}\left[
                  1+\ep^2\frac{360\zeta(3)}{\pi^3}(RT)^3+\dots
                  \right]\,
\ee
for the free energy and
\be\label{f3} f^{\rm PFA}=-\frac{\pi^3 R}{360 d^3}\left[
                 1+\ep^38(RT)^4+\dots\right]\,
\ee
for the force. The other special case is $RT\gg1$. It corresponds
to medium temperature and we come to
\be\label{F4}
                  \F^{\rm PFA}=-\frac{\pi^3 R}{720 d^2}\left[
                  1+\ep^220(RT)^2+\dots
                  \right]\,
\ee
for the free energy and
\be\label{f4} f^{\rm PFA}=-\frac{\pi^3 R}{360 d^3}\left[
                 1+\ep^3\frac{360\zeta(3)}{\pi^3}(RT)^3+\dots\right]\,
\ee
for the force. In \Ref{F4} we used that the integral
$\int_0^\infty dt\, t^{-3} g(t)=5/2$ could be calculated
explicitly.

Now we consider   medium and high temperature as defined in
\Ref{4.tr}. For this we substitute
\be \label{lmh} RT=\frac{dT}{\ep} \ee
in \Ref{F1} and \Ref{f1} and consider the limit $\ep\to0$ with
$dT$ fixed. In this limit, \Ref{F1} and \Ref{f1} turn into
\be\label{F5}
                  \F^{\rm PFA}=-\frac{\pi^3 R}{720 d^2}\left[
                  1+h(dT)+\dots
                  \right]\,
\ee
for the free energy and
\be\label{f5} f^{\rm PFA}=-\frac{\pi^3 R}{360 d^3}\left[
                 1+g(2 dT)+\dots\right]\,
\ee
for the force. In \Ref{F5} we defined
\be\label{h}h(x)=2\int_1^\infty dt\,\frac{g(2tx)}{t^3}. \ee
The limiting values of this function are
\be\label{has} h(x)\sim    \left\{
        \begin{array}{c}20\,x^2 +\dots \quad\mbox{for $x\to0$,}\\[5pt]
        \frac{180\zeta(3)}{\pi^3}\,x -1+\dots \quad\mbox{for $x\to\infty$,}
        \end{array}  \right.
\ee
where the dots denote exponentially small contributions.  We will
meet this function again in section 4. We mention that \Ref{4.f1}
holds for the contributions displayed in \Ref{F5} and \Ref{f5},
see Eq.\Ref{5.hd}.

The expansions \Ref{F5} and \Ref{f5} are uniform in $dT$. For
$dT\ll1$ these turn just into \Ref{F4} and \Ref{f4},
correspondingly. In this way, at medium temperature, the
asymptotics of  \Ref{F4} and \Ref{f4} for $RT\to\infty$ just match
the asymptotics of \Ref{F5} and \Ref{f5} for $dT\to0$. For
$dT\gg1$, i.e., for high temperature, useing the lower line in \Ref{has}
 the asymptotics are
\be\label{F6}
\F^{\rm PFA}=
-\frac{\zeta(3)RT}{4d}+\dots
\ee
for the free energy and
\be\label{f6}
f^{\rm PFA}=
-\frac{\zeta(3)RT}{4d^2}+\dots
\ee
for the force. Here the
dots denote exponentially fast decreasing contributions.

In this way, with equations \Ref{F2} and \Ref{f2} we derived the
free energy and the force in PFA for small and medium temperature.
For fixed $RT$ these are small, suppressed by 2 resp. 3 orders of
$\ep$. Obviously, the rule \Ref{rule} does not hold. The square
bracket in \Ref{f2} should be equal to the parenthesis in
\Ref{4.F} taken for small $dT$ but as seen from \Ref{g3} it is
different. This is in agreement with the general discussion
following the rule \Ref{rule} since for small $dT$ the thermal
contribution to $\F_{||}$ has $\alpha\le0$. It is only in the
limiting case, for $RT\gg 1$, Eq.\Ref{f4}, that \Ref{rule} holds.

For medium and large temperature the free energy and the force are
given by Eqs.\Ref{F5} and \Ref{f5}. In this case the thermal
corrections are not suppressed by powers of $\ep$. Their seize
depends on $dT$. For small $dT$ they are small and match \Ref{F2}
and \Ref{f2}  taken at large $RT$. For $dT\sim1$ they are of the
same  order as the vacuum energy and for $dT\gg1$ they dominate.
For the force \Ref{f5}, the rule \Ref{rule} holds since in this
case one has to take the function $g(x)$ for large argument in the
parenthesis in \Ref{4.F}. This is also in agreement with the
general discussion following the rule \Ref{rule} since this
behavior corresponds to an $\alpha>0$. In this way, the rule
\Ref{rule} can be generalized to finite $T$ in the region of
medium and large temperature but not at small temperature. It must
be mentioned that at small temperature the temperature dependent
part is very small and therefore not relevant in applications.

\section{The free energy at small separation}
In this section we consider the exact expression \Ref{1.F2} for
the free energy at small separation, i.e., for $\ep\ll 1$. First
we assume fixed values of $R$ and $T$. In the sense of the
definition \Ref{4.tr} this is the region of  low and medium
temperature. Also we restrict ourselves to the case of a scalar
field with Dirichlet boundary conditions. We make use of a number
of notations and the methods developed in
\cite{Bordag:2006vc,BORDAG2008C}. We start with expanding the
logarithm in \Ref{1.F2}, i.e., with the representation
\be\label{2.F1}
\F=-\frac{T}{2}\sum_{n=-\infty}^{\infty}
\sum_{s=0}^{\infty}\frac{1}{s+1}\,
\sum_{m=-\infty}^{\infty}   \sum_{l=|m|}^{\infty}
\left(  \prod_{j=1}^{s}\sum_{n_j=|m|-l}^{\infty}  \right)    {\cal Z}(\xi_n),
\ee
where we defined
\be\label{2.Z}
{\cal Z(\xi)}=\prod_{i=0}^{s}M_{l+n_i,l+n_{i+1}}(\xi)
\ee
for the product of the (s+1) matrices $\mathbf{M}(\xi)$. We adopt
the formal setting $n_0=n_{s+1}=0$. Up to the frequency sum, this
representation is identical to the corresponding one in
\cite{BORDAG2008C}. For instance,  the convergence properties are
the same. For any finite $\ep$, all sums in \Ref{2.F1} converge.
However, for $\ep$ becoming  smaller, a higher and higher number
of terms give significant contributions until in the limit
$\ep\to0$ the convergence gets lost.

In \cite{BORDAG2008C}, the behavior for $\ep\to0$ was established
by making an asymptotic expansion. First, the orbital momentum
sums were  substituted by corresponding integrations, afterwards a
substitution of variables allowed for performing the limit. We
will use that later in this section but first we  proceed in a
slightly different way. We first transform the Matsubara sum into
integrations using the Abel-Plana formula. In this way we come
from \Ref{2.F1} to
\be\label{2.F2} \F=E_0+\F_T
\ee
with
\be\label{2.FT} \F_T=-
\frac{1}{2}
\sum_{s=0}^{\infty}\frac{1}{s+1}\,
\int_{-\infty}^\infty\frac{d\xi}{2\pi}\ n_T(\xi)
 \sum_{m=-\infty}^{\infty}
\sum_{l=|m|}^{\infty} \left(
\prod_{j=1}^{s}\sum_{n_j=|m|-l}^{\infty}  \right) i\left( {\cal
Z}(i\xi)- {\cal Z}(-i\xi)\right). \ee
The first term in the rhs. of \Ref{2.F2} is just the vacuum energy \Ref{1.E}
which appeared from the direct substitution of the Matsubara sum by the
integration over $\xi$. The second contribution, $\F_T$,  involves
the Boltzmann factor,
\be\label{2.n}   n_T(\xi)=\frac{1}{e^{|\xi|/T}-1},
\ee
and it can be viewed as the 'pure temperature contribution'. The
analytic continuation $\xi\to i\xi$ is, so to say, the return to
real frequencies. From the structure of the integrand one can show
that there are no singularities the integration path could cross during rotation.
Especially there are no singularities on the real frequency axis.
If there were some these would correspond to discrete eigenvalues
of the Laplace operator in the region between the sphere and the
plane. However, since that is an open geometry, the spectrum is
pure continuous.

Now, for the vacuum energy $E_0$  we can use the known asymptotic expansion for
$\ep\to0$. In leading order it is the PFA.
Including the first correction beyond PFA it was calculated in \cite{BORDAG2008C} for the
scalar field and in \cite{BORDAG2009C} for the electromagnetic field.
For example, for Dirichlet boundary conditions on the sphere and on the plane it reads
\be\label{2.E2}
E_0=-\frac{\zeta(4)}{16\pi R}\,\frac{1}{\ep^2}\left(1+\frac{1}{3}\,\ep+\dots\right).
\ee
For the other boundary conditions  this formula
holds too, only the coefficients change. For the electromagnetic field logarithmic
contributions appear.

Now we consider the pure temperature contribution  \Ref{2.FT}. In
order to perform the analytic continuation $\xi\to\pm i\xi$, we
use the corresponding analytic continuations of the Bessel
function entering \Ref{1.M} and \Ref{1.d}. These are (see, for
example   \cite{AbramowitzStegun})
\bea\label{2.acont1} I_\nu(iz)=i^\nu J_\nu(z), &~~& K_\nu(i
z)=\frac{-i\pi}{2}\, i^{-\nu}H^{(2)}_\nu(z),
\\
I_\nu(-iz)=i^{-\nu} J_\nu(z), && K_\nu(-i z)=\frac{i\pi}{2}\,
i^{\nu}H^{(1)}_\nu(z).
 \eea
We insert these into the matrix $\mathbf{M}(\xi)$, \Ref{1.M}, and
obtain
\be\label{2.M1} M^{\rm  }_{l,l'}(i \xi) =\frac{J_{\nu'}(\xi
R)}{H^{(2)}_\nu(\xi R)} \sqrt{\frac{\pi}{4\xi
L}}\sum_{l''=|l-l'|}^{l+l'}(-1)^{(l+l'-l'')/2}
H^{(2)}_{\nu''}(2\xi L)H_{ll'}^{l''}\,, \ee
where we took into account that only $l''$ contribute for which
$(l+l'-l'')$ is an even number. Next we use
$H^{(1,2)}_\nu(z)=J_\nu(z) \pm i Y_\nu(z)$ in order to separate
the imaginary part,
\bea\label{2.M2}
M^{\rm  }_{l,l'}(i \xi)
&=&
\frac{J_{\nu'}(\xi R)}{Y_\nu(\xi R)}
\sqrt{\frac{\pi}{4\xi L}}\sum_{l''=|l-l'|}^{l+l'}(-1)^{(l+l'-l'')/2}
Y_{\nu''}(2\xi L)H_{ll'}^{l''}\,
\frac{1+r_\nu(\xi R)r_{\nu''}(2\xi L)}{1+\left(r_\nu(\xi R)\right)^2}
\nn\\ &&  \times
\left[1-i\frac{ r_\nu(\xi R)-r_{\nu''}(2\xi L)}{1+r_\nu(\xi R)r_{\nu''}(2\xi L)}\right]
 \,,
\eea
where we defined the ratio
\be\label{2.ratio}r_\nu(\xi)=\frac{J_\nu(\xi)}{Y_\nu(\xi)}.
\ee
In fact we need the imaginary part of ${\cal Z}(i\xi)$, i.e., of
the product of (s+1) matrices \Ref{2.M2}. We observe, that the
imaginary part of this product always contains at least one factor
$r_\nu(\xi)$.

Now we consider the convergence properties of the temperature
dependent part  \Ref{2.FT}. In opposite to the zero temperature
case, the convergence of the integration over $\xi$ is now
provided by the Boltzmann factor. We are left with the question
about the convergence of the orbital momentum sums. It is known,
that these cease to converge for $\ep\to0$ in the vacuum energy
where the modified Bessel functions enter. In the temperature
dependent part we have the 'non-modified' Bessel functions. We
need their asymptotic expansion for large indexes. Again using
\cite{AbramowitzStegun} we note
\be\label{2.asJY}
J_\nu(z)=\sqrt{\frac{1}{2\pi\nu}}\,e^{-\nu\ln\nu+1+\nu\ln(z/2)}
                      \left(1+O\left(\frac{1}{\nu}\right)\right),\qquad
Y_\nu(z)=\sqrt{\frac{2}{\pi\nu}}\,e^{\nu\ln\nu-1-\nu\ln(z/2)}
                      \left(1+O\left(\frac{1}{\nu}\right)\right).
\ee
From here it follows that the ratio $r_\nu(\xi)$ is fast
decreasing for increasing index $\nu$. As a consequence, the
imaginary part of $\cal Z$ entering $\F_T$, makes the orbital
momentum sums  convergent irrespective of $\ep$. We note that this
is in opposite to the real part where the convergence comes from
the combination $J_{\nu'}(\xi R)Y_{\nu''}(2\xi L)/Y_\nu(\xi R)$
and is present only if $R<L$, or equivalently, $\ep>0$, hold.

In this way, the integral and the sums in \Ref{2.FT} remain finite
if we put $\ep=0$ directly  in $\cal Z$. In this way, $\F_T$ has a
finite limit for $\ep\to0$, $RT$ fixed. We mention that the same
result was obtained in the preceding  section using PFA.
Therefore, from the representation of the free energy used in this
section, which is an exact one, we conclude that
 the temperature dependent part of
the free energy is at small separation a next-to-next-to-leading
order correction. The leading order is given by the vacuum energy.
It is proportional to $1/\ep^2$. After that there come the beyond
leading order corrections proportional to $1/\ep$ to the vacuum energy which were
calculated analytically in  \cite{Bordag:2006vc,BORDAG2008C,BORDAG2009C}.
Only after that, in order
$\ep^0$, the temperature corrections start to contribute.

\begin{figure}[h]
\includegraphics{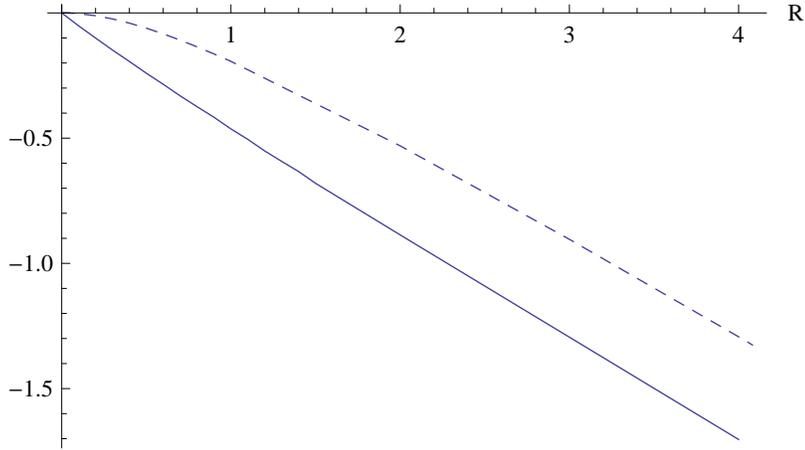}
\caption{The temperature dependent part $\F_T$ of the free energy
(in arbitrary units)
of a scalar field for $\ep=0$ and $T=1$ as
function of $R$ calculated from the exact formula  (solid line)
and the corresponding proximity force approximation (dashed
line).\label{figAP1}}
\end{figure}

\begin{figure}[h]
\includegraphics{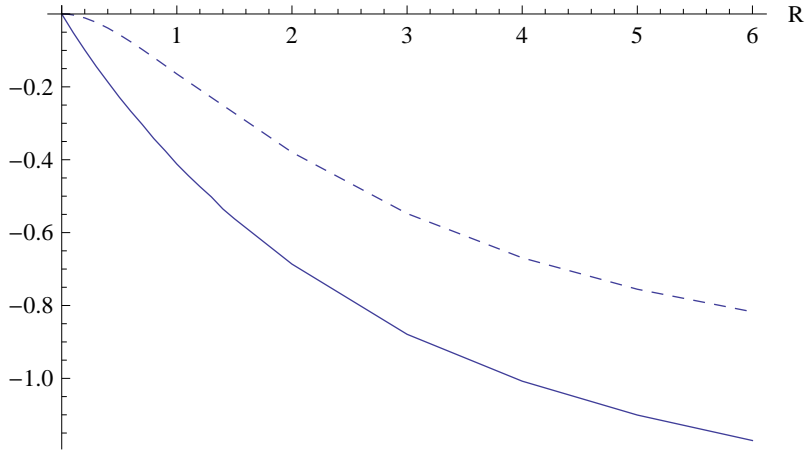}
\caption{The temperature dependent part $\F_T$ of the free energy
(in arbitrary units)
of a scalar field  for $\ep=0.1$ and $T=1$ as
function of $R$ calculated from the exact formula  (solid line)
and the corresponding proximity force approximation (dashed
line).\label{figAP2}}
\end{figure}

The above discussion holds for $R$ and $T$ kept fixed and $d$
small. This covers the region of low temperature until the border
between low and medium temperature with $RT\sim 1$. In this region
a numerical evaluation of the temperature dependent part is
possible. We have made this calculation using the representation
\be\label{3.FTt} \F_T= \frac{T}{2\pi}\int_{0}^\infty d\xi \
n_1(\xi) \, i \,\Tr \left[\ln (1-\mathbf{M}(i\xi T))-\ln
(1-\mathbf{M}(-i\xi T))\right], \ee
which is the same as Eq.\Ref{2.FT} but without expanding the
logarithm and after the substitution $\xi\to \xi T$. The results
are represented in Fig. \ref{figAP1} and \ref{figAP2} for $T=1$ as
function of $R$. Fig. \ref{figAP1} is for $\ep=0$ and Fig.
\ref{figAP2} for $\ep=0.1$. For small $R$ the calculation is quite
easy since only the lowest orbital momenta contribute. In the
limit of   $R\to0$ the temperature part $\F_T$ is the same as
obtained from the low temperature expansion which will be
considered in the next section. For increasing $R$, more and more
orbital momenta need to be included. In the process of evaluation
we increased this number until the relative change dropped below
$10^{-2}$. For example, for $R=2$ we had to include orbital
momenta until $l=9$.

An interesting feature of this calculation is that it allows for a
direct comparison with the PFA results derived in the preceding
section. The corresponding values for $\F_T^{\rm PFA}$ calculated
from \Ref{4.FP2} are shown in the Fig. \ref{figAP1} and
\ref{figAP2} as dashed lines. It can be clearly seen that the PFA
deviates from the true values of $\F_T$ quite significantly. This
statement holds for small temperature and at least up to the
border between low and medium temperatures. For example, in Fig.
\ref{figAP2} we cover the region from $R=0$ until $R\sim 6$ which
by means of the definition \Ref{4.tr} is the beginning of the
region of medium temperature. At $R=6$, the PFA result for $\F_T$
is only about $70\%$ of the true value. The picture becomes better
for PFA if calculating the temperature dependent contribution to
the force by means of \Ref{4.f1}. We made a calculation for
several values of $\ep$ and $R$ with $T=1$. The results are
displayed in Tables \ref{tab1} and \ref{tab2}.
\begin{table}[t]\caption{The temperature dependent part of the force from the
exact formula and from the PFA for $\ep=0.01$ and
$T=1$.\label{tab1}}
\begin{ruledtabular}
\begin{tabular}{ccccc}
 &$R$&0.5&1&3\\\hline
&$f_T$ & 0.14&0.59&5.1 \\
&$f_T^{\rm PFA}$  & 0.047&0.33&4.6
\end{tabular}
\end{ruledtabular}\end{table}
\begin{table}[t]\caption{The temperature dependent part of the force from the
exact formula and from the PFA for $\ep=0.1$ and
$T=1$.\label{tab2}}
\begin{ruledtabular}
\begin{tabular}{ccccc}
&$R$&0.5&1&6\\
&$f_T$ & 0.14&0.56&7.9 \\
&$f_T^{\rm PFA}$  & 0.047&0.31&8.2
\end{tabular}
\end{ruledtabular}\end{table}
It is seen that for increasing $R$ the deviation of the PFA value
from the exact one decreases much faster than that for the energy.
Already for $R=6$, $\ep=0.1$, it amounts only a few percent. This
allows for the conclusion that at low temperature the PFA does not
give correct results. However, for increasing temperature, the
deviation of PFA from the exact results gets smaller and already
at the border between low and medium temperatures the PFA gives
quite good results for the temperature dependent part of the
force.


Now we consider the region of medium and high temperature by
keeping $d$ and $T$ fixed while $\ep\to0$. We start from
representation \Ref{1.F2} of the free energy for the
electromagnetic field. We rewrite the Matsubara sum by inserting
unity in the form
\be\label{5.Eins}1=\int_{-\infty}^\infty d\xi\,\delta(\xi-\xi_n).
\ee
Interchanging the orders of summation and integration we come to
the representation
\be\label{5.F3a} \F=\frac{T}{2}\int_{-\infty}^\infty d\xi\,
    \sum_{n=-\infty}^{\infty} \delta(\xi-\xi_n)\ \Tr \ln
\left(1-\mathbf{M}(\xi)\right). \ee
Now we use the relation
\be\label{5.Po}
\sum_{n=-\infty}^\infty \delta(\xi-\xi_n)
=\frac{1}{2\pi T}\sum_{n=-\infty}^\infty e^{i n\xi/T},
\ee
which is equivalent to the Poisson resummation formula, and the
free energy takes the form
\be\label{5.F4a} \F=\frac{1}{2}\int_{-\infty}^\infty
\frac{d\xi}{2\pi}\,
    \sum_{n=-\infty}^{\infty} e^{i n\xi/T}
    \ \Tr \ln \left(1-\mathbf{M}(\xi)\right).
\ee
This expression differs from the vacuum energy \Ref{1.E} only by
the sum and the additional exponential factor.
For this reason we can apply the same methods as  in
\cite{BORDAG2008C,BORDAG2009C} to calculate the asymptotic
behavior for $\ep\to0$. First we expand the logarithm and come to
a representation in parallel  to \Ref{2.F1},
\be\label{5.F1} \F=-\frac{1}{2R}
\int_{-\infty}^\infty\frac{d\xi}{2\pi}\,
\sum_{n=-\infty}^{\infty}   e^{i n\xi/(RT)}
\sum_{s=0}^{\infty}\frac{1}{s+1}\, \sum_{m=-\infty}^{\infty}
\sum_{l=|m|}^{\infty} \left(
\prod_{j=1}^{s}\sum_{n_j=|m|-l}^{\infty}  \right)    {\cal
Z}\left(\frac{\xi}{R}\right), \ee
where $ {\cal Z}$ is given by Eq.\Ref{2.Z}. In \Ref{5.F1}, in
addition we substituted $\xi\to\xi/R$. After that, $Z(\xi/R)$ is a
function of $\ep$ only. Now we consider the limit $\ep\to 0$. For
decreasing $\ep$, the main contribution to the integral over $\xi$
and and to the sums over the orbital momenta come from higher and
higher values. Therefore we substitute the orbital momentum sums
by corresponding integrals and change the variables according to
\bea\label{5.28} \xi=\frac{t}{\ep}\,\sqrt{1-\tau^2}, \quad
l=\frac{t}{\ep}\,\tau, \quad m=\sqrt{\frac{t\tau}{\ep}}\, \mu\,,
\nn \\
\quad  n_i=\sqrt{\frac{4t}{\ep}}\,\tilde{n}_i~~(i=1,\dots,s)\,.
\eea
In this way we get an asymptotic expansion in the form
\bea\label{5.F2} \F&=& -\frac{R}{4\pi d^2} \sum_{s=0}^\infty
    \frac{1}{s+1} \int_0^\infty dt \, t  \ e^{-2t(s+1)}
    \int_{0}^1 \frac{d\tau\,{\tau}}{\sqrt{1-\tau^2}}\,
    \sum_{n=-\infty}^{\infty}   e^{i nt\sqrt{1-\tau^2}/(\ep RT)}
 \nn\\&& \times
     \int_{-\infty}^\infty {d\mu} \ e^{-\mu^2(s+1)}\
    \left(\prod_{j=1}^s \int_{-\infty}^\infty\frac{d\tilde{n}_j}{\sqrt{\pi}} \right)
  \cal{Z}^{\rm as}\,,
\eea
where
\be\label{5.Zas} {\cal Z}^{\rm as}=
    \prod_{i=0}^s   \left( \sqrt{\frac{4\pi t}{\ep}}\,
    M^{\rm }_{l+n_i,l+n_{i+1}}\right)\,e^{2t(s+1)+\eta_1+\mu^2 (s+1)}\,,
 \ee
with $\eta_1=\sum_{i=0}^s(\tilde{n}_i-\tilde{n}_{i+1})^2$. We stress again that
these formulas are in complete analogy to the corresponding ones
in \cite{BORDAG2008C,BORDAG2009C}. The whole difference is in the
sum over $n$ and the T-dependent exponential. For instance, the
factor ${\cal Z}^{\rm as}$ is the same as without temperature.
Therefore we know it has an expansion
\be\label{5.Zas1}{\cal Z}^{\rm as}=1+O(\ep).\ee
The corrections $O(\ep)$ depend only on the geometry and the kind
of fields, but not on the temperature. Now we are interested in
the pure temperature dependent contributions. Hence we drop these
corrections and take ${\cal Z}^{\rm as}=1$. After that $\F$,
Eq.\Ref{5.F2}, can be simplified. First of all, one can perform the
integrations over $\tilde{n}_i$ and $\mu$,
\bea\label{5.F3} \F&=& -\frac{R}{4\pi d^2} \sum_{s=0}^\infty
    \frac{1}{(s+1)^2} \int_0^\infty dt \, t  \ e^{-2t(s+1)}
    \int_{0}^1 \frac{d\tau\,{\tau}}{\sqrt{1-\tau^2}}\,
    \sum_{n=-\infty}^{\infty}   e^{i nt\sqrt{1-\tau^2}/(\ep RT)}.
\eea
Now the integration over $t$ is trivial. Also the integration over
$\tau$ can be carried out leaving  the double sum
\bea\label{5.F4} \F&=& -\frac{R}{4\pi d^2} \sum_{s=0}^\infty
    \frac{1}{(s+1)^2}
    \sum_{n=-\infty}^{\infty}  \frac{\ep^2 T^2}{n^2+4\ep^2(s+1)^2T^2}.
\eea
Here the contribution from $n=0$ is just the zero temperature contribution, i.e., the vacuum energy.
For the remaining sum over $n\ne0$ we use the formula
\be\label{5.sum}\sum_{n=1}^\infty\frac{1}{a^2+n^2}=\frac{a\pi\coth(a\pi)-1}{2a^2}
\ee
and the free energy can be written in the form
\bea\label{5.F5} \F&=& -\frac{R}{4\pi d^2}\left(1+h(2dT)\right),
\eea
where we defined
\be\label{5.h} h(x)=90 x^4\sum_{m=1}^\infty \left[\frac{\coth(m\pi
x)}{(m\pi  x)^3}-\frac{1}{(m\pi  x)^4}\right]. \ee
In the last step we substituted $s+1\to m$.

In \Ref{5.F5}, the function $h(2dT)$ describes the temperature
contribution to the free energy for medium and high temperature at
small separation. It was derived from the exact formula
\Ref{1.F2}. Now it is interesting to observe that this function,
$h(x)$ as given by Eq.\Ref{5.h}, is just the same as the function
$h(x)$, Eq.\Ref{h}, found in PFA in the preceding section. This
can be shown, for example, by inserting the sum representation
\Ref{g2} for $g(x)$ into \Ref{h}. For this it is useful to rewrite
\Ref{g2} in the form
\be\label{5.g}g(x)=45x^4\sum_{m=1}^\infty
\frac{1}{y}\frac{\pa}{\pa
y}\left(-\frac{\coth(y)}{y}+\frac{1}{y^2}\right)_{\big|{y=m\pi
x}}. \ee
Obviously, the integration in \Ref{h} can be carried out
explicitly and one comes just to Eq.\Ref{5.h}. In this way, we
confirm the PFA \Ref{F5} and with it all subsequent discussions in
section 3.

A similar statement holds for the force. In taking the derivative
with respect to the separation $d$ of \Ref{5.F5}, we have to
consider the derivative of $h(x)/x^2$. For this derivative the
following formula holds,
\be\label{5.hd}\frac{\pa}{\pa
x}\frac{h(x)}{x}=-\frac{2}{x^3}\,g(x), \ee
where $g(x)$ is the same function as in the temperature dependent
part of the free energy for parallel plates \Ref{4.F}. This can be
seen directly by inserting \Ref{5.h} into the left side of
\Ref{5.hd} and comparing with \Ref{g2}. In this way, also
Eq.\Ref{f2} for the force in PFA is shown to coincide with the
exact expression obtained from the T-matrix representation
\Ref{1.F2}.

Finally we mention that the function $h(x)$, like $g(x)$,
Eq.\Ref{inversion}, has an inversion symmetry,
\be\label{5.inversion}h\left(\frac{1}{x}\right)=\frac{5}{x^2}-\frac{h(x)}{x^4},
\ee
which can be derived from \Ref{inversion} and \Ref{5.hd}.

\section{The free energy at low temperature}
In this section we consider the low temperature expansion of  the
free energy, i.e., we assume $T\ll 1/R$ and $T\ll 1/d$ not
restricting the relation between $d$ and $R$. We start from
representation \Ref{2.F2}. The lowest approximation for $T\to0$
is, of course, the vacuum energy. The temperature corrections come
from expanding $\F_T$, \Ref{2.FT}, for small $T$. In this section
it is useful not to expand the logarithm and we take the
representation
\be\label{3.FT}
\F_T=
\frac{1}{2\pi}\int_{0}^\infty{d\xi}\ n_T(\xi)
\, i \, \Tr  \left[\ln(1-\mathbf{M}(i\xi))-\ln(1-\mathbf{M}(-i\xi))\right].
\ee
For $T\to0$, because of the Boltzmann factor, we can expand the
remaining part of the integrand in \Ref{3.FT} in powers of $\xi$.
Each additional power in $\xi$ will add a power in $T$. So we need
the lowest odd power of $\xi$. The lowest even power of $\xi$ does
not contribute to the difference between the logarithms which in
fact represent the jump across the cut the logarithm has.

First we expand $M_{l,l'}(\xi)$, \Ref{1.M}. We use the ascending
series of the modified Bessel functions (see, for example
\cite{AbramowitzStegun}),
\be\label{3.IK}
I_\nu(z)=\left(\frac{z}{2}\right)^\nu\frac{i_\nu}{\Gamma(\nu+1)},
\qquad
K_\nu(z)=(-1)^l\,\frac{\pi}{2}\left(\frac{z}{2}\right)^{-\nu}
\left(\frac{i_{-\nu}}{\Gamma(1-\nu)}-
\left(\frac{z}{2}\right)^{2l+1}\frac{i_{\nu}}{\Gamma(1+\nu)}\right),
 \ee
where $i_\nu$ has only even powers of $z$ and we continue to use
the convention $\nu\equiv l+1/2$. Inserting these expansions into \Ref{1.M}
and remembering that only $l''$ give non-zero contributions for which $l+l'-l''$ is even,
we observe that the first odd power of $\xi$ comes from the
second term in $K_\nu(z)$ in \Ref{3.IK}. Moreover, the lowest odd power comes from the lowest orbital momenta
$l$ and $l'$ only. For this reason from the orbital momentum sums only one term remains
and the logarithm can be calculated directly. For Dirichlet resp. Neumann boundary conditions on the sphere we
obtain in this way for the scalar field
\bea\label{3.M0}
M_{0,0}^{\rm D}(\xi)&=&\frac{R}{2L}-\left(1-\frac{R}{2L}\right)R\xi+\dots\,,
\nn\\
M_{0,0}^{\rm N}(\xi)&=&-\frac{R^3}{6L}\xi^2+\frac{1}{3}R^3\xi^3+\dots\,.
\eea
The logarithms are
\bea\label{3.M1}
\ln\left(1-M_{0,0}^{\rm D}(\xi)\right)&=&\ln\left(1-\frac{R}{2L}\right)+R\xi+\dots
\nn \\
\ln\left(1-M_{0,0}^{\rm N}(\xi)\right)&=& \frac{R^3}{6L}\xi^2-\frac{1}{3}R^3\xi^3+\dots
\eea
and the jumps become
\bea\label{3.M3}
i \, \Tr  \left[\ln(1-\mathbf{M^{\rm D}}(i\xi))-\ln(1-\mathbf{M^{\rm D}}(-i\xi))\right]
=-2R\xi+\dots\,,
\nn \\
i \, \Tr  \left[\ln(1-\mathbf{M^{\rm N}}(i\xi))-\ln(1-\mathbf{M^{\rm N}}(-i\xi))\right]
=-\frac{2}{3}R^3\xi^3+\dots\,.
\eea
Finally we have to insert this into \Ref{3.FT}. The remaining
integration can be done in terms of zeta functions and we come to
\bea\label{3.FTDX}
\F_T^{\rm D}&=&-\frac{\zeta(2)}{\pi}\,RT^2+\dots\,,
\nn \\
\F_T^{\rm N}&=&-\frac{2\zeta(4)}{\pi}\,R^3T^4+\dots\,.
\eea
These formulas are for Dirichlet boundary conditions on the plane.
For Neumann boundary conditions on the plane we have to reverse the sign in the logarithms
 in \Ref{3.FT}. A simple calculation in parallel to the above one results in
\bea\label{3.FTNX}
\F_T^{\rm D}&=&\frac{\zeta(2)}{\pi}\frac{2L-R}{2L+R}\,RT^2+\dots\,,
\nn \\
\F_T^{\rm N}&=&\frac{2\zeta(4)}{ \pi}\,R^3T^4+\dots\,.
\eea
We see that the leading order in the low temperature expansion
follows the behavior for low momentum of the corresponding
T-matrix. For Dirichlet boundary conditions, from the s-wave, the
lowest order is $\xi$ resulting in correction of order  $T^2$. For
Neumann boundary conditions, due to the derivatives, this
contribution is absent and the expansion starts with $T^4$.

For the electromagnetic field we have to take care of the
polarizations. However, the off-diagonal entries of the matrix
$\mathbf{M}$ start with an additional  power in $\xi$ resulting
from $\tilde{\Lambda}$. Since these enter the trace in quadratic
combinations only, an additional power of $\xi^2$ as compared to
the diagonal contributions  results. As a consequence, the
off-diagonal entries do not contribute to the leading order at
small $T$. So we are left with, separately, the contributions from
the TE and TM modes. For these, the same considerations as for the
scalar field apply with the only difference of \Ref{1.dTM} in
place of \Ref{1.d} and the orbital momentum sum starting from the
p-wave. We write down explicitly the relevant contributions to
$M_{l,l'}(\xi)$ for both modes and for relevant values of the
azimuthal index $m$,
\bea\label{3.m1}
M_{1,1}^{{\rm TE},m=0}(\xi)&=&
\frac{R^3}{8L^3}-\frac{R^3}{4L}\left(1-\frac{3R^2}{10L^2}\right)\xi^2
                 +\frac{R^3}{3}\left(1-\frac{R^3}{8L^3}\right)\xi^3+\dots\,,
 \nn\\
M_{1,1}^{{\rm TE},m=1}(\xi)&=&
\frac{R^3}{16L^3}+\frac{R^3}{8L}\left(1+\frac{3R^2}{10L^2}\right)\xi^2
                 -\frac{R^3}{3}\left(1+\frac{R^3}{16L^3}\right)\xi^3+\dots\,,
\nn\\
M_{1,1}^{{\rm TM},m=0}(\xi)&=&
\frac{R^3}{4L^3}-\frac{R^3}{2L}\left(1+\frac{3R^2}{20L^2}\right)\xi^2
                 +\frac{2R^3}{3}\left(1+\frac{R^3}{4L^3}\right)\xi^3+\dots\,,
 \nn\\
M_{1,1}^{{\rm TM},m=1}(\xi)&=&
\frac{R^3}{8L^3}+\frac{R^3}{4L}\left(1-\frac{3R^2}{20L^2}\right)\xi^2
                 -\frac{2R^3}{3}\left(1-\frac{R^3}{8L^3}\right)\xi^3+\dots\,.
\eea
From here we calculate the jumps just like in \Ref{3.M3}.
Introducing the symbolic notation $\Delta$ for these we note
\bea\label{3.m2}
\Delta^{{\rm TE},m=0}&=&-\frac{2}{3}\,R^3\xi^3+\dots\,,
\nn \\
\Delta^{{\rm TE},m=1}&=&\frac{2}{3}\frac{16L^3+R^3}{16L^3-R^3}\,R^3\xi^3+\dots \,,
\nn\\
\Delta^{{\rm TM},m=0}&=&-\frac{4}{3}\frac{4L^3+R^3}{4L^3-R^3}\,R^3\xi^3+\dots \,,
\nn\\
\Delta^{{\rm TM},m=1}&=&\frac{4}{3}\,R^3\xi^3+\dots \,.
\eea
These expressions we have to insert into $\F_T$ and to integrate
over $\xi$. In the end we get
\bea\label{3.FTED}
\F_T=\frac{6\zeta(4)}{\pi}
\left(1+\frac{2}{3}\frac{16L^3+R^3}{16L^3-R^3}-\frac{2}{3}\frac{4L^3+R^3}{4L^3-R^3}\right)
\,R^3T^4+\dots\,.
\eea
This is the low temperature correction for the electromagnetic
field. We see that it gives an order $T^4$ contribution like for
the forces acting between parallel plates.

Now we can consider \Ref{3.FTED} for small separation. We can put
$d=0$ directly and get
\be\label{3.T0}\F_T=\frac{58\zeta(4)}{15\pi}\,R^3T^4+\dots\,. \ee
This expression is, as expected, different from the temperature
corrections in PFA, Eq.\Ref{F3}. It gives the low temperature
(this is $T\ll 1/R$) correction beyond PFA.

It is also possible to consider \Ref{3.FTED} for large separation,
or equivalently, for small $R$. One gets
\be \label{3.To}
\F_T=\frac{6\zeta(4)}{\pi}\left(1-\frac{R^3}{4d^3}+\dots\right)\,R^3T^4\,.
\ee
The leading order coincides with the corresponding low temperature
expansion of Eq.(4) in \cite{Durand2009}.
%
\section{The free energy at high temperature}
The high temperature expansion can best be analyzed starting from
the original Matsubara sum \Ref{1.F2}. The leading order for
$T\to\infty$ is given by the contribution with $n=0$, i.e., by the
lowest Matsubara frequency.

We separate   the contribution from $n=0$,
\be\label{4.F1}
\F=T F_0\left(\frac{d}{R}\right)+F_1(Td,TR)
\ee
and   the  contributions with $n\ne0$ are collected in $F_1$. The
leading contribution $F_0$ is proportional  to $T$ and it is a
function of the dimensionless ratio \Ref{2.d}.  The function $F_1$
depends on two dimensionless combinations.

We consider the function $F_0$ in more detail. It is given by the
formula
\be\label{4.F0}
F_0(\ep)=\frac{1}{2}\,\Tr \ln (1-\mathbf{M}(0)).
\ee
From \Ref{1.M} and using  \Ref{3.IK} we get for the scalar field
with Dirichlet boundary conditions on the sphere
\be\label{4.MD}
M_{l,l'}^{\rm D}(0)=\left(\frac{R}{2L}\right)^{l+l'+1}
\frac{\sqrt{\pi}\, \Gamma(l+l'+1/2)}{2\Gamma(l+1/2)\Gamma(l'+3/2)}\,
H_{l,l'}^{l+l'}.
\ee
Here we  took into account that from the sum over $l''$ in
\Ref{1.M} only the term with $l''=l+l'$ did contribute. For Neumann
boundary condition on the sphere we have to account for the
derivatives. These result in a simple factor and we can write
\be\label{4.MN}
M_{l,l'}^{\rm N}(0)=\frac{-l'}{l+1}\,M_{l,l'}^{\rm D}(0)
\ee
For the electromagnetic field we have in addition the terms mixing
the polarizations. However, since $\tilde{\Lambda}_{l,l'}$ is
proportional to $\xi$ these do not contribute. So we are left with
the additional factor $\Lambda_{l,l'}^{l''}$ and the difference in the derivatives.
We come to
\be\label{4.MTX}
M_{l,l'}^{\rm TE}(0)= M_{l,l'}^{\rm D}(0) \,\Lambda_{l,l'}^{l+l'},
\qquad
M_{l,l'}^{\rm TM}(0)= \frac{l'+1}{l}\,M_{l,l'}^{\rm D}(0) \,\Lambda_{l,l'}^{l+l'}.
\ee
These expressions can be inserted into \Ref{4.F0} and for any finite $\ep=d/R\ne 0$
the function $F_0$ can be calculated numerically.

We consider the limiting cases of $\ep\to\infty$ (large
separation) and $\ep\to0$ (small separation). For large
separation we simply have to expand the $M_{l,l'}(0)$ into powers
of $1/\ep$ and see that only the lowest orbital momenta
contribute. We obtain
\bea\label{4.58}
M_{0,0}^{\rm D}(0)= \frac{R}{2L}+\dots,
&~~~&
M_{0,0}^{\rm N}(0)=-2\left(\frac{R}{2L}\right)^3+\dots,
\nn\\
M_{1,1'}^{\rm TE}(0)= 2\left(\frac{R}{2L}\right)^3+\dots,
&&
M_{1,1}^{\rm TM}(0)= 4\left(\frac{R}{2L}\right)^3+\dots\,.
\eea
The corresponding functions $F_0$ we get from \Ref{4.F0} in this
approximation by multiplication with $(-1/2)$. For the
electromagnetic field, i.e., adding the two contributions in the
lower line in \Ref{4.58}, the result coincides with the lower line
in Eq.(6) in \cite{Durand2009}.

In the opposite limit of small separation, i.e., for $\ep\to0$ we
are faced with the problem that the convergence of the orbital
momentum sums gets lost. This problem is essentially the same as
at zero temperature in the same limit.  Even more, it can be
treated by the same methods, i.e., by calculating the asymptotic
expansion of $F_0(\ep)$ for $\ep\to0$. The first step in this
procedure is to expand the logarithm in \Ref{4.F0} and to
substitute the orbital momentum sums by corresponding
integrations,
\be\label{4.F02}
F_0(\ep)=
-\sum_{s=0}^{\infty}\frac{1}{s+1}
\int_0^\infty dl\,     \int_{-l}^{l}dm
\, \left(\prod_{j=1}^s\int_{l-m}^\infty dn_j\right)  {\cal Z}
\ee
with
\be\label{4.Z}{\cal Z}=\prod_{i=0}^sM_{l+n_i,l+n_{i+1}}(0)\,.
\ee
We mention that this is, up to the sums substituted by integrals,
the $(n=0)$-contribution in \Ref{2.F1}. Next we have to make an
appropriate substitution of variables,
\be\label{4.subst} l=\frac{t}{\ep},\quad
m=\sqrt{\frac{t}{\ep}}\,\mu,\quad
n_i=\sqrt{\frac{4t}{\ep}}\,\tilde{n}_i\,.
\ee
This is the same substitution as used in \cite{BORDAG2008C},
Eq.(8) and in \cite{BORDAG2009C}, Eq.(28), here however restricted
to $\tau=1$. Therefore we can use the asymptotic formulas derived
there. For instance, in leading order we have
\be\label{4.M1}
M_{l+n,l+n'}(0)=\left(\frac{\ep}{4\pi t}\right)^{1/2}\ e^{-2t-(n-n')^2-\mu^2}
(1+O(\ep)).
\ee
This expression is the same for both boundary conditions and also
for the modes of the electromagnetic field. For the latter there is in this
approximation also no mixing of the polarizations, for details see \cite{BORDAG2009C}.
Making in \Ref{4.F02} the substitution \Ref{4.subst} we come for the electromagnetic field  with \Ref{4.M1} to
\be\label{4.F03} F_0(\ep)= -\frac{1}{2\ep}
\sum_{s=0}^{\infty}\frac{1}{s+1} \int_0^\infty dt\,e^{-2t(s+1)}
\int_{-\infty}^\infty\frac{d\mu}{\sqrt{\pi}}\ e^{-\mu^2(s+1)}
\left(\prod_{j=1}^s\int_{-\infty}^\infty
\frac{d\tilde{n}_j}{\sqrt{\pi}}\right) e^{-\eta_1}+\dots \ee
with
$\label{4.eta1}\eta_1=\sum_{i=0}^s(\tilde{n}_i-\tilde{n}_{i+1})^2.
$
Carrying out the integrations and the sum we obtain
\be\label{4.F04}F_0(\ep)=-\frac{\zeta(3)}{4\ep}+\dots
\ee
We mention that this formula  coincides with the corresponding PFA
result \Ref{F6} although we did not assume $d\ll R$.

In general, the contribution of the zeroth Matsubara frequency is
equivalent to a theory with a dimension reduced by one. However,
thereby it is usually assumed that the original theory (taken in
its Euclidean version)  has a symmetry between the spatial and the
time dimensions. One must bear in mind that in the given case this
symmetry is broken by the boundaries.

Finally we consider $F_1$ in \Ref{4.F1}, i.e., the contributions
from the non-zero Matsubara frequencies. Here we observe   for
large $T$ an exponential decrease of the functions $M_{l,l'}(0)$,
\be\label{4.F11}
M_{l,l'}(0)\sim e^{-4\pi n T d}
\ee
($d=L-R$) such that $F_1$ is exponentially suppressed for all
$d>0$. This is the same property as we observe in  the case of
parallel planes. This suppression holds for for $\frac{1}{T}\ll
d$, $\frac{1}{T}\ll R$. If in addition $d\ll R$, i.e.
$\frac{1}{T}\ll d\ll R$, holds this suppression disappears. In
that case, as before, higher and higher $n$ and  orbital momenta
contribute. In fact, this is the limit $RT\to\infty$ in the region
of high temperature as defined in section 3, Eq.\Ref{4.tr}. In
that case it is appropriate use representation \Ref{5.F4a} and to
proceed as it was done in the second part of section 4. As a
result, $F_1$ would deliver a correction beyond what is displayed
in Eq.\Ref{5.F5}.


\section{Conclusions}
In the foregoing sections we calculated the free energy for a
sphere in front of a plane at finite temperature for conductor
boundary conditions. First we considered the case of small
separation. Here we used both methods, the PFA and the functional
determinant representation and considered both, the free energy
and   the force. In the region of low temperature, the thermal
contributions to the free energy and to the force are very small
and PFA does not hold for them. In the region of  medium and high
temperatures, which is the temperature region of experimental
interest, we have reproduced the PFA from the exact method. For
instance we have shown that the rule \Ref{rule} does hold in this
case.

We would like to mention that in  \cite{Milton:2009gk} exact
finite temperature results were obtained for the interaction
between a plane and a semitransparent curved surface described by
a delta-potential. For weak coupling it was shown that the free
energy of a scalar field coincides at all temperatures with the
proximity force approximation corresponding to this geometry.

Next we considered low temperature without restriction to small
separation and derived the corresponding expansions for the free
energy. Here the behavior of the thermal contribution to the free
energy depends on the boundary conditions and on the kind of the
field. An interesting question on the interplay between
temperature and geometry was raised in
\cite{Weber:2009dp,Gies:2009nn}. There it was observed that in an
open geometry at low temperature lower powers of the temperature
may appear as compared with the parallel planes. In general, we
cannot support this conclusion from our calculations. Only for a
scalar field with Dirichlet boundary conditions we observe
$\F_T\sim T^2$, Eq.\Ref{3.FTDX}. For Neumann boundary conditions
and for the electromagnetic field we see $\F_T\sim T^4$,
Eqs.\Ref{3.FTDX} (lower line), \Ref{3.FTNX} and \Ref{3.T0}.

In section 6 we considered the limit of high temperature. As
expected, here the dimensional reduction works and the leading
order contribution comes from the zeroth Matsubara frequency. At
small separation, the asymptotics could be calculated using the
same methods as for zero temperature and the result coincides with
that of PFA.

 \vspace{1cm}\noindent This work was supported
by the Heisenberg-Landau program. The authors acknowledge helpful
discussions with G.Klimchitskaya and V.Mostepanenko. The authors
benefited from exchange of ideas by the ESF Research Network
CASIMIR.

\bibliographystyle{unsrt}\bibliography{C:/Users/bordag/WORK/Literatur/articoli,C:/Users/bordag/WORK/Literatur/libri,C:/Users/bordag/WORK/Literatur/Bordag}

\end{document}